\documentstyle[11pt,aaspp4,tighten]{article}
\def\rosat{{\sl ROSAT~}}
\def\mpc{{ $h_{50}^{-1}$~Mpc}}
\slugcomment{}
\begin{document}

\title{\bf A2125 and its Environs:\\
Evidence for an X-ray-emitting Hierarchical Superstructure}
\author{Q. Daniel Wang}
\affil{Dearborn Observatory, Northwestern University}
\affil{ 2131 Sheridan Road, Evanston,~IL 60208-2900}
\affil{E-mail: wqd@nwu.edu}

\affil{Andrew Connolly\altaffilmark{1} and Robert Brunner\altaffilmark{1}}
\affil{Department of Physics and Astronomy}
\affil{Johns Hopkins University, Baltimore, MD 21218}
\affil{ajc@tiamat.pha.jhu.edu and rbrunner@hoplite.pha.jhu.edu}

\altaffiltext{1}{Guest observer at the Kitt Peak National
Observatory, National Optical Astronomy Observatories, which is
operated by the Association of Universities for Research in Astronomy,
Inc. (AURA) under cooperative agreement with the National Science
Foundation.}

\begin{abstract}

Based on a deep {\sl ROSAT}/PSPC observation, we reveal an elongated
complex of extended X-ray-emitting objects in and around the galaxy
cluster A2125. Multicolor optical imaging of galaxies in the field
suggests that this complex represents a hierarchical superstructure
spanning $\sim 11$\mpc\ at redshift $\sim 0.247$. The
multi-peak X-ray morphology of A2125 suggests that the cluster is an
ongoing coalescence of at least three major subunits.  The dynamical
youth of this cluster is consistent with its large fraction of blue galaxies
observed by Butcher \& Oemler.  The superstructure contains two
additional clusters, projected at distances of only 3 and 4.3\mpc\ 
from A2125. But the most interesting feature is the low-surface-brightness 
X-ray emission from a moderate galaxy concentration not associated with 
individual clusters. The emission
likely arises in a hot ($\sim 10^7$~K) intergalactic medium, as
predicted in N-body/hydro simulations of structure formation. 

\end{abstract}

\section {Introduction}
	
	A substantial fraction, probably most, of the baryonic matter
in the Universe is expected to be in a hot phase of the intergalactic medium 
(IGM). This hot IGM is a definite prediction of recent N-body/hydro simulations
of structure formation (e.g., Ostriker \& Cen 1996). During the structure 
formation, shocks induced by the collisions between large-scale
collapsing flows can naturally
heat the IGM to a temperature range of $\sim 10^5-10^7$~K. 
The integrated emission from the hot IGM (Cen et al. 1995) provides a
reasonably good explanation of the diffuse extragalactic $\sim 0.7$~keV 
background inferred from the Magellanic Bridge X-ray shadowing 
experiment (Wang \& Ye 1996). The simulations further show that the hot IGM
forms filamentary superstructures (cosmic webs or caustics), on scales 
greater than individual clusters, and that the IGM emission is
greatly enhanced in regions near clusters, where gravitational
potential wells are relatively deep. Analytical modeling for the triaxial 
collapse of cosmological perturbations also predicts the presence of
X-ray-emitting filaments near clusters (Elsenstein \& Loeb 1995). 
Properties (e.g., temperature, chemical abundance, axial ratio) of such 
filaments can provide important constraints on theories of galaxy feedback 
and cluster formation.

	Here, we report the detection of such an X-ray-emitting 
superstructure. Along with our systematic X-ray study of Butcher 
\& Oemler clusters (1984), we find that A2125, an optically 
rich cluster of a high blue galaxy fraction ($f_b = 19\pm3\%$) at $z = 0.247$,
is associated with a complex of galaxies, clusters
and diffuse hot gas. Extending $\sim 11$\mpc, ($h_{50}$ --- 
the Hubble constant in units of $50 {\rm~km~s^{-1}~Mpc^{-1}}$; $q_o=0.1$ 
is assumed throughout the paper), this hierarchical complex closely resembles 
filamentary superstructures seen in the simulations (e.g., Fig. 10 in 
Bryan et al. 1994; Bryan 1996, personal communications). 

\section {Observations and Data Reduction}

	The X-ray observation (rp800511) of A2125 was made with the 
\rosat PSPC (Positional Sensitive
Proportional Counter; Tr\"umper 1992 and references therein). The
effective exposure is 19359~s. We
processed the data, following the standard procedure for studying
diffuse X-ray sources (Snowden et al. 1994). We limited our imaging
analysis in the 0.5-2~keV band (PSPC channels 13-30) to reduce both
the foreground from the Galactic hot interstellar medium, which
dominates below $\sim 0.3$~keV, and the background variation caused by
a differential foreground X-ray absorption in the field. The spatial
variation in the Galactic X-ray-absorbing gas column density, inferred
from the {\sl IRAS} 100$\mu$m survey, is up to about $2 \times 10^{20}
{\rm~cm^{-2}}$ in the field and should result in a differential
absorption $\lesssim 10\%$ in the 0.5-2~keV image. Furthermore, our
analysis was based on data within $16^\prime$ of the 
PSPC pointing axis. Outside this central field, the analysis might
otherwise be complicated by both the detector's window supporting
structure (at $\sim 20^\prime$ off-axis) and the degraded point spread
function (PSF; Hasinger et al. 1994).

	We classified X-ray sources of signal-to-noise ratios greater
than 3 into two classes, point-like and extended (Fig. 1), based on
maximum likelihood fits to the 2-D count distributions of individual
X-ray sources. Extended sources are those with their count
distributions broader than the off-axis dependent PSF at confidences
$\gtrsim 95\%$. Some of these sources may represent peaks of large-scale
diffuse X-ray structures (e.g., clusters). 
We excised point-like sources by replacing counts within 85\% source
flux-encircled radii with randomly generated events of intensities
interpolated from neighboring averages. The residual flux of the
excised sources, spread over the image due to the very extended wing
of the PSF, produces no discernible local enhancement in Fig. 2.

As a follow-up of this X-ray study, we obtained multicolor CCD
images covering the extended X-ray features in the field. Using the
Prime Focus camera on the Kitt Peak Mayall 4-meter telescope, we
observed A2125 and the extended X-ray sources 
to the north and to the southwest of the cluster. We also obtained
a pointing just beyond the southwestern corner of Fig. 2.. All 
these pointings were 
undertaken in the U, B, R and I passbands. With a pixel scale of 0\farcs47, 
each T2KB CCD pointing covered a 16\arcmin\ field of view 
(equivalent to 4.9\mpc\ at $z=0.247$).

	We reduced these optical data following standard procedures in
the IRAF image reduction packages. We detected objects using the
SExtractor package (Bertin and Arnouts 1996) with a detection kernel
of 1\farcs2 (equivalent to the FWHM of the seeing disk). Aperture and
Kron magnitudes were measured for all detected sources. Colors of
these sources were derived using a 20\arcsec\ diameter aperture. The
detection limits for these photometric data were 22.7, 22.5, 23.5 and
22.4 in the U, B, R and I passbands respectively (for a 5 sigma
detection of a point source). All magnitudes were converted to the
Johnson photometric system using Landolt standards.

\section {Analyses and Results}

	Fig. 2 shows a complex of extended X-ray-emitting
objects. Possibly limited by the size of the field, this complex spans
at least 35\arcmin\ (from the upper middle to the lower right). Fig. 3
presents the apparent associations of the X-ray-emitting objects with
concentrations of galaxies. The optically known rich cluster A2125
(Abell richness 4) appears surprisingly irregular in the X-ray, and is
detected as a combination of six extended X-ray peaks. The two new
clusters (Clusters B and C hereafter) are evidently the optical
counterparts of the two extended X-ray objects (Figs. 3b,c) north of
A2125.  There are three more extended X-ray peaks to the south and west
of A2125 within the 16\arcmin\ off-axis (Fig. 1); their association with
galaxy concentrations are less certain.  The double peaks 
at $\sim 15^h39^m57, 66^\circ6\arcmin$ has an extended radio
counterpart, apparently in the 1.4 GHz NRAO Sky VLA survey. The radio
and X-ray emissions may thus represent the interaction between a jet
from an active nucleus and its ambient medium (e.g., Doe et al. 1995).
We cannot, however, exclude the possibility that such an apparently extended 
peak actually results from the confusion between two or more
point-like sources. Besides the clusters and peaks, the complex
contains substantial amounts of low-surface brightness X-ray emission
(LSBXE).  The emission is most prominent in the region southwest to
A2125 within a position angle between $\sim 200^\circ$ and $250^\circ$
(anti-clockwise from North), relative to the centroid of the cluster. The
emission between $\sim -20^\circ$ and $20^\circ$ may also be
considerable, but is difficult to quantify because of the confusion
with discrete sources.  We concentrate on characterizing the three
most prominent components: A2125, Cluster B, and the southwestern
LSBXE, which are all coincident with over-densities in the galaxy
distributions.

Table 1 lists X-ray spectral properties of the individual components, including
the centers and sizes of the regions from which individual spectra are
extracted. A background spectrum is estimated in a partial annulus
between 8\arcmin\ and 16\arcmin\ centered on A2125, excluding the two
pie-like regions of the LSBXE mentioned above. We fit the spectra with the 
standard Raymond \& Smith thermal plasma model, plus a foreground absorption 
with an assumed metal abundance of the X-ray-absorbing gas as
100\% solar. The fitted absorptions ($N_H$) toward the two clusters are
consistent with the 21~cm measured Galactic neutral hydrogen 
column density of $2.9 \times 10^{20} {\rm~cm^{-2}}$ of the field (Stark et
al. 1992). We fix the abundance of the ICM as 20\% solar. Varying the 
abundance within its expected uncertainty of
a factor 2 (Mushotzky et al. 1996) changes the obtained spectral 
parameters by less than 10\%. The abundance of the X-ray-emitting gas
responsible for the LSBXE component is unknown. Our best-fit is 2\% solar, 
but no meaningful 90\% confidence limits are obtained. The component 
also provides little constraints on the absorption. We thus fix the 
X-ray absorbing-gas column density as the 21~cm measured value. 

	From the best-fit temperature $kT \sim 0.85$~keV of the LSBXE
component, one can estimate the hot gas emission measure  as $EM \sim (3.2
\times 10^{-2} {\rm~cm^{-6}~pc}) (S_o/10^{-4}
{\rm~counts~s^{-1}~arcmin^{-2}}$), using the observed surface
brightness $S_o$ (Fig. 2). The mean gas density is then $\sim (1.8
\times 10^{-4} {\rm~cm^{-3}}) (S_o/10^{-4}
{\rm~counts~s^{-1}~arcmin^{-2}})^{1/2} (l_{Mpc})^{-1/2}$, where
$l_{Mpc}$ is the effective depth of the gas in units of
Mpc. If the X-ray-emitting gas fills uniformly a cylinder of 
diameter 2.5\mpc\ (8\arcmin) and length 3.4\mpc\ (11\arcmin), for example,
the luminosity in Table 1 then suggests a total gas mass $M_g \sim 7 \times
10^{13} M_\odot$. 
 
	We characterize the azimuthally-averaged surface brightness
profiles of the two clusters, using the standard $\beta$ model
(Cavaliere \& Fusco-Femiano 1976). A fit of the model 
(including a background parameter) to the profile of Cluster B
($\chi^2/d.o.f. = 29.6/36$) gives the $\beta$ value, the core radius $r_c$
and the central electron density $n_{e,c}$ as 
1.3(0.86-4.5), 0.65(0.43-1.5)\mpc, and $1.1(0.88-1.3) \times 10^{-3} 
h_{50}^{1/2}{\rm~cm^{-3}}$ (90\% confidence limits). The two parameters
$\beta$ and $r_c$ are strongly correlated with each other. We have 
also assumed an approximately spherical and isothermal state of the ICM
in deriving $n_{e,c}$. The total ICM mass within an 1.5\mpc radius
is $\sim 5.4\times 10^{13} M_\odot$.
 A direct $\beta$ model fit to the A2125 profile
($\chi^2/d.o.f = 57.6/36$) can be ruled out at the 99\% confidence.
The deviation of the profile from the model is primarily due
to the presence of the three
distinct peaks in the central region of the cluster (Fig. 2). To improve the
fit we remove the data within a circular region around the centroid of 
each peak. The radius of this region is chosen to be the 85\%
source-flux-encircled radius of a point-like source.
The fit to the peak-excised profile is acceptable ($\chi^2/d.o.f. = 44.3/36$). 
The obtained $\beta$ model parameters are
$\beta = 0.77(0.62-1.1),\ r_c = 0.52(0.37-0.76)$\mpc, and 
$n_{e,c} = 1.0(0.83-1.2) \times 10^{-3} h_{50}^{1/2}{\rm~cm^{-3}}$. 
The estimated ICM mass is $8.4 \times 10^{13} M_\odot$ 
within the 1.5\mpc\ radius.

We further characterize the 2-D X-ray morphology of A2125 and
Cluster B. We first fit the X-ray morphology of each cluster with a series 
of ellipses, using iteratively-calculated moments of the PSPC count 
distribution (Buote \& Canizares 1994; Wang, Ulmer, \& Lavery 
1997). Cluster B is statistically consistent
with being circular; the ellipticity is $\lesssim 0.31$ (95\%
confidence). A2125 has an ellipticity of 0.38(0.29-0.48) and a North-to-East
position angle of 132(124-140)~deg, on scales
between 2\arcmin-6\arcmin, and shows multi-peaks on
smaller scales. We then calculate the the peak-excised surface brightness 
profile of the cluster as a function of 
the semi-major axis in the best-fit elliptical coordinates. The $\beta$ model
fit to the profile is satisfactory ($\chi^2/d.o.f. = 31.3/36$), and yields
the model parameters as $\beta = 0.77(0.59-1.1),\ r_c = 0.51(0.33-0.84)$\mpc, 
and $n_{e,c} = 1.0(0.80-1.3) \times 10^{-3} h_{50}^{1/2}{\rm~cm^{-3}}$ 
(assuming an oblate shape of the ICM).

	We follow an approach similar to that described by Neumann \&
B\"ohringer (1996) to reveal the substructure of A2125.  We first
subtract the best-fit elliptical $\beta$ model ($S_m$) 
from the data ($S_o$), and then smooth the residual map by convolving it with 
a Gaussian ($g$), i.e., $(S_o-S_m) \oplus g$.  We further calculate the
noise map as [max$(S_o/t \oplus g^2$, $S_m/t \oplus g^2)]^{1/2}$,
where $t$ is the exposure map.  The  residual-to-noise ratio
gives a significance measure of surface
brightness residuals. Fig. 4 shows both the residual and significance maps. 
The three central peaks, standing out with their
residual-to-noise ratios greater than 3, are of comparable sizes $\sim
10\arcsec\pm4\arcsec$ (after the PSF contribution is subtracted in
quadrature) and luminosities $\sim 1.6-2.3 \times 10^{43} h_{50}^{-2}
{\rm~ergs~s^{-1}}$.

	The association of the X-ray components with the galaxy
over-densities are evident in Fig. 3. While we are yet to make a
spectroscopic redshift mapping of galaxies in the field, we can use
the colors of the galaxies to determine whether the E/S0 sequences in
each of the components are consistent with their being at the same
redshift. For an elliptical galaxy at a redshift of 0.247 the Johnson
$B-R$ color is $\sim 2.5$ (Fukugita et al. 1995). We find that A2125
and the Cluster B have distinct E/S0 sequences in their $B-R$ vs $R$
color magnitude diagram. The $B-R$ color of these sequences is 2.54
and 2.45 for A2215 and cluster $B$ respectively. Given the uncertainty
of approximately 10\% in the colors, the two systems are consistent
with a redshift of 0.247. Furthermore, if we isolate those galaxies
with $2.4 < B-R < 2.6$, we find that the LSBXE correlates with an
over-density of galaxies. We thus tentatively identify the complex as
a coherent superstructure, which spans $\sim 11$\mpc\ at
the redshift $z = 0.247$. The moderate galaxy over-density
appears to extend further to the north, beyond the 
field covered by Figs. 1-2, but not to the southwest. We will present
more detailed analysis of the optical observations later.

\section {Discussion}

	The X-ray-emitting complex associated with A2125 is morphologically 
very similar to hierarchical superstructures seen in various
N-body/hydrodynamic simulations (e.g., CDM$+\Lambda$ universe; Cen \&
Ostriker 1994; one may find movies of the simulations at
http://dept.physics.upenn.edu/bode/TVD/). Such
superstructures represent caustics where the initial density
fluctuation peaks and large-scale collapsing gas flows intersect.  In
addition to the shock-heating of the gas, both the thermal and
chemical feedback from stars may also play an important role. 
The heating and chemical enrichment of the IGM affect our
understanding the structure and 
evolution of clusters (e.g., Kaiser 1991; David, Jones,\& Forman 1996). 
The hot IGM may also be responsible for the positive
cross-correlation between Abell clusters and the X-ray background
surface brightness at $\sim 1$~keV, as detected by Soltan et
al. (1996).

	Our detection of the LSBXE in the vicinity of A2125 provides
probably the first direct observational evidence for hot gas outside
individual clusters. Our measured temperature of $kT \sim
0.85$~keV is at the high end of the range predicted for the hot IGM.
This is natural because the X-ray emission, proportional to the square
of gas density, traces deep gravitational potential valleys in the
Universe.  Because no bright nearby early-type galaxy is present in
the field, the LSBXE cannot be due to the intragroup hot gas of a
nearby group of galaxies.  Most likely, both the hot gas and the galaxy
concentration represent part of a hierarchical superstructure that
includes Clusters B and C, and A2125 as well. The scale and morphology
of both the hot gas and galaxy distributions indicate that this
superstructure is not in a relaxed equilibrium state.  The entire
superstructure can, however, collapse within a few $10^9$ years to
form a single cluster.
 
	A2125 itself also appears to be a young 
system.  First, the presence of at least three major X-ray
subcomponents indicates that the A2125 is a  coalescence of subunits 
of comparable masses. Such a coalescent occurs typically only at the initial 
stage of rich cluster formation. Second, this coalescence scenario explains 
both the strong elongation of A2125 and its 
gross misalignment with the orientation of the overall
superstructure. Mergers of a rich cluster with small subclusters, taking place
at later stages of cluster evolution, tend to  
produce a cluster elongation along the superstructure. The measured ellipticity
of A2125 is the largest in a sample of 10 Butcher \& Oemler 
clusters observed with the PSPC (Wang \& Ulmer 1997 in preparation).
Third, the relatively low temperature and luminosity of A2125 (Table 1), 
compared to relaxed clusters of similar optical
richness (e.g., Edge \& Steward 1992), indicate that the ICM has not
yet virialized to the gravitational potential of the system.  Fourth,
the high fraction of blue galaxies observed in A2125 can be a
natural consequence of the coalescence of spiral-rich poor clusters
(Kauffmann 1995; Schindler \&
Wambsganss 1996; Wang, Ulmer, \& Lavery 1997). Therefore, A2125, as a rich
assembly of galaxies, is still at its early evolutionary 
stage to become a rich X-ray-emitting cluster.
	
	Further observations of the
A2125 field can yield more quantitative information on the IGM,
galaxies, and dark matter in the superstructure. Ongoing optical photometry
and spectroscopic surveys of the field will provide information on
both the galaxy population and dynamics. The field is also
ideal for probing the large-scale mass distribution by observing
weakly lensed background galaxies, and for measuring the hot gas
column density by mapping out microwave
background distortions through the Sunyaev-Zel'dovich effect (Persi et
al. 1995).

\acknowledgements
	
	We thank G. Bryan and J. Ostriker for valuable comments on the
work, and acknowledge the support by NASA under grants NAG-5-2716,
NAG-5-2717 (QDW), and AR-06394 (AJC) as well as the Graduate Student 
Researchers Program (RJB).
\clearpage
\begin{table}[htb]
%\scriptsize
\begin{tabular}{lccc} 
\multicolumn{4}{c}{\bf Table 1: Spectral Properties 
of the Three Major Components\tablenotemark{a}}\\ [0.1in]
\hline\hline 
Parameter 	& A2125		&Cluster B	&LSBXE\\
\hline      
R.A. (J2000)    &  15  41 06.3  & 15 41 10.6 	&15  40  2\\
Dec. (J2000)    &  +66 16 10    & +66  26 26	& +66  10 0 \\
Size (arcmin)	&  5.0		&2.5		& $\sim 8\times$11\\
Count rate ($10^{-3} {\rm~counts~s^{-1}}$)	&42.9 	&20.6 & 18.9\\
$\chi^2/d.o.f$	&27.4/28	&8.6/18		&4.8/9	\\
$kT$ (keV) & 2.3(1.6-4.5)	&1.6(1.2-2.8)   &0.85 (0.46-1.3)\\
 $N_H (10^{20} {\rm~cm^{-2}})$	&2.5(1.7-3.4)	& 3.6(2.5-6.0) &2.9 (fixed) \\
Luminosity ($10^{43} h_{50}^{-2}{\rm~ergs~s^{-1}}$)	&15	& 6.4 &4.0 \\
\hline 
\end{tabular}
\tablenotetext{a}{Uncertainties in parameter values,
as presented in parentheses, are at the 90\% confidence level.
The luminosities are estimated with the
best-fit spectral parameters and in the $z=0.247$ rest-frame 
energy range of 0.5-2~keV.
}
\end{table}

%\clearpage
\clearpage
\begin{figure} \caption{PSPC image of A2125 and its vicinity in the 0.5-2~keV 
band. The image, corrected for exposure, is smoothed adaptively with 
a Gaussian, the size of which is adjusted at each pixel to achieve a 
count-to-noise ratio of 4. Each contour is 50\% 
(2$\sigma$) above its lower level; the lowest contour
is at $1.8 \times 10^{-4} {\rm~counts~s^{-1}~arcmin^{-2}}$. 
Detected X-ray source positions are represented with {\sl crosses}; 
those extended ones are further marked with {\sl squares}.
\label{fig1}}
\end{figure}

\begin{figure} \caption{The image is the same as in Fig. 1, except 
it is smoothed after point-like sources are excised. 
\label{fig2}}
\end{figure}

\begin{figure} \caption{Overlays of the diffuse X-ray 
surface brightness contours (Fig. 2)
on blue-band CCD images covering A2125 and the LSBXE region (a),
Cluster B (b), and Cluster C (c). 
\label{fig3}}
\end{figure}

\begin{figure} \caption{X-ray substructures of A2125. The upper panel shows
the PSPC surface brightness residuals after the best-fit 
elliptical $\beta$ model is subtracted; the contours are at -6.8, -2.8, 
5.2, 9.2, 17.2, and 33.2 $\times 10^{-3} {\rm~counts~s^{-1}~arcmin^{-2}}$. 
The lower panel presents the significance of the residuals, measured 
with the residual-to-noise ratio; the contours are at
-2, -1, 1, 2, 3, 4, 5$\sigma$. Negative contours outline regions of possible
over-subtraction of the model.
\label{fig4}}
\end{figure}

\end{document}